# Electronic structure and stability of hydrogen defects in diamond and boron doped diamond: A density functional theory study

Ashutosh Upadhyay[1*], Akhilesh Kumar Singh[1], Amit Kumar[1,2,*]

[1]School of Materials Science and Technology, Indian Institute of Technology (Banaras Hindu University), Varanasi-221005, India

[2]Institut Néel, CNRS and Université Joseph Fourier, 25 rue des Martyrs, BP166, 38042 Grenoble cedex 9, France

## Abstract

Isolated hydrogen and hydrogen pairs in bulk diamond matrix have been studied using density functional theory calculations. The electronic structure and stability of isolated and paired hydrogen defects are investigated at different possible lattice sites in pure diamond and boron doped diamond. Calculations revealed that isolated hydrogen defect is stable at bond center sites for pure diamond and bond center puckered site for boron doped diamond. In case of hydrogen pairs, $H_2^*$ defect (one hydrogen at bond center and second at anti-bonding site) is stable for pure diamond, while for boron doped diamond B-$H_{2BC}$ complex (one H atom at the B-C bond centered puckered position and the other one at the puckered position of one of the C-C bond first neighbor of the B atom) is most stable. Multiple hydrogen trapping sites in boron doped diamond has also been studied. Calculated results are discussed and compared with previously reported theoretical results in detailed.

[*]**Corresponding author mail ID:** amitnsc@gmail.com, u.ashutosh@yahoo.com

## I. INTRODUCTION

Diamond is a wide band gap semiconductor with exceptional physical and electronic properties [1-4]. Recent advances in the synthesis of single crystal diamond via chemical vapor deposition (CVD) have demonstrated that material with exceptional electronic and optical properties can now be produced, and this has enlivened interest in exploiting the extraordinary properties of diamond [5]. In particular, diamond exhibits very high electron (4500 $cm^{-2}$ $V^{-1}$ $s^{-1}$) and hole (3800 $cm^{-2}$ $V^{-1}$ $s^{-1}$) mobilities [5], breakdown strength ($10^7$ V $cm^{-1}$) and thermal conductivity (>2000 W $m^{-1}$ $K^{-1}$) [6], which enable diamond to surpass other wide band gap materials for high power and high frequency electronic applications. However, the electronic properties of diamond are altered by defects and impurities grown into the material and introduced during the processing steps in device fabrication. Therefore, it needed to understand the electronic properties and lattice structures of defects or impurities in diamond matrix.

Hydrogen related defects play an important role in diamond and strongly affects the electronic and structural properties. Hydrogen is usually the most abundant element present in the CVD growth environment of diamond and is used in many processing techniques. It is now established that under the appropriate conditions hydrogen can be readily incorporated into bulk diamond. Detailed information on the structure of hydrogen-related defects in diamond is an essential prerequisite for understanding the influence of hydrogen on electrical and optical properties, and ultimately exploiting the full potential of these materials. Furthermore, diamond is the ideal model wide band gap material for both experimental and theoretical investigations of hydrogen-intrinsic defect complexes.

Isolated hydrogen atom and hydrogen atom pairs in diamond matrix have been attracted researchers over the last decades. In last couples of years, the energies or stabilities of several sites for isolated hydrogen in diamond have been theoretically calculated using different theoretical approaches. Most of the theoretical calculations [7-10] predicted the bond center sites as a most stable geometry, although one semiempirical calculations suggested an off-axis sites to be more stable [11]. For H atom pairs, different geometries are proposed and reported in literature. Two nearby H atoms has been predicted to be more stable, when one carbon atom sits on bond center site and other on the same axis but on the other side of one bond center carbon atoms, which is called the anti-bonding site. The H atoms pairs is called $H_2^*$ defect. This geometry of H atom pairs is found highly unstable in Si and Ge semiconductors. Silverman et al. reported that the average hydrogen interstitial formation energy in the amorphous region is lower than the hydrogen interstitial formation energy in nano-diamond [12].

Despite the basic interest of isolated H atom defect, H atom has been shown to neutralize the electrical activity of both n-type and p-type dopants and to produce a variety of defects and impurity states. This has been studied extensively more than twenty years in Si and Ge semiconductors [13-14]. In recently years, it has been noted that hydrogen can also passivate impurities in diamond [15-16]. In last few years, there has been intense theoretical activity on boron-hydrogen complex in diamond matrix [17-22]. The formation of boron-hydrogen or boron-deuterium complexes convert p-type boron doped sample into highly insulating sample as revealed by capacitance-voltage measurements [15] and Hall measurement [23-24]. Other spectroscopic techniques Fourier transform Infrared [25] and cathodolumiensecence measurements also confirm the passive nature of hydrogenated B-doped diamond [26]. It is also reported that excess hydrogen or deuterium plasma exposed B-doped layers shows the n-type conductivity with shallow activation energy ($\approx 0.23$ eV) [27-28], which has been very interesting topic in last few years [29]. This has been a controversial topic in theoretical calculations. Some density functional calculations support the experimental findings [20-21], whereas other calculations rule out the proposed explanation of n-type conductivity by the formation of boron-hydrogen (B-H) complex formation [17-19].

In this paper, we report on the electronic structure and stability for the isolated H atom and H atom pairs defects in pure diamond and boron doped diamond are studied by means of density functional theory (DFT) using plane wave method. Different possible geometries are studied and a comprehensive view of isolated H defect and H atom pairs defects is reported. The main objective of this work is to present detailed ab initio calculations of the total energy, structure optimization and densities of states (DOS) for the H related defect diamond matrix in different possible geometries. All the calculations have been done by using supercell technique. To reduce the possible errors due to calculation parameters we have used the large cell and large set of k-points. Calculation details are discussed in brief in section II, Theoretical calculation for hydrogen defect in pure diamond are given in section III and H defect in B-doped diamond are described in section IV.

## II. CALCULATION DETAILS

The electronic structure and stability of isolated hydrogen and hydrogen atom pairs complexes were investigated using the Vienna Ab-initio Simulation Package (VASP) code [30]. It is based on DFT within the generalized gradient approximation (GGA) [31]. Projector Augmented wave is used with a basis cutoff equal to 318.6 eV [32]. The Monkhorst-Pack scheme [33] with 5x5x5 k points has been used for integration in the Brillouin zone. The error of total energy convergence is less than $10^{-4}$ eV. Tetrahedron method with Bloechl corrections have been used to calculate the densities of states. The calculation of the energy band gap for pure diamond (E$g$ = 4.2 eV) is in good agreement with theoretical calculations [9, 11, 34]. To generate a puckered position, the hydrogen atom was initially displaced by a small distance off-axis and the whole cell was allowed to relax.

Fig. 1 shows the schematic of the selected sites of hydrogen atoms in diamond cell. Carbon atoms are shown by gold circles as a diamond cell. The hydrogen atoms at different interstitial sites are shown by black circles. The Tetrahedral (T) site lies equidistant from four carbon sites and possesses T$_d$ symmetry, the Hexagonal (H) site lies midway between two T sites and possesses D$_3d$ symmetry. The bond-centered (BC) site is the mid-point between two atom sites (D$_3d$ symmetry) and anti-bonded (AB) configuration is opposite the BC site along the same axis, possesses C$_3\upsilon$ symmetry [13].

## III. HYDROGEN IN PURE DIAMOND

### A. Isolated Hydrogen in pure diamond

For isolated H atom, there are mainly three different sites namely; bond-center (BC), Tetrahedral (T), and Hexagonal (H) site. We performed the calculations for all the three possible sites of isolated hydrogen in diamond matrix. Fig. 2 shows the relaxed structures of hydrogen at (a) Bond-centered, (b) Tetrahedral (T) and (c) Hexagonal site. Total energy calculations show that the bond-centers site is the most stable one. For this configuration H atom sits at equidistant from both carbon atoms with 1.14 Å C-H bond lengths. The observed C-H bond length for H at bond center is higher than the previously reported value. The higher bond length could be possible due to full relaxation and use of bigger size supercell in present study. In the present study, the bond lengths and bond angles are mentioned within the error bar of ± 0.01 angstrom and ±0.1 degree, respectively.

Some earlier calculations using the ab initio Hartree-Fock (HF) approach reported the tetrahedral interstitial site (T) is a deep minimum of the total energy [35-37]. Bond-centered (BC) site was also found more stable compare to the tetrahedral interstitial site (T) using basis-set ab initio HF [7, 38, 39] and later by DFT calculations [8]. In early stage, Estreicher et al. [37] predicted the T site as a most stable but later in fully relaxed cluster, they also reported BC site as a most stable site for isolated hydrogen. Tachikawas also discussed the electronic states of hydrogen atom trapped in diamond using cluster model calculations and predicted that the tetrahedral site is most favored site for H trapping in diamond matrix. He calculated the C-C bond length was slightly elongated by the Insertion of Hydrogen atom in diamond lattice [40]. A very different site off-bond axis was also reported to be the lowest energy position that an H atom occupies in diamond. It is a six fold degenerate site with respect to the C-C bond, which is denoted by "Equilateral Triangle" ET site [11]. This site was not reproduced by the others. Single H at anti-bonding (AB) site was also calculated, which is highly unstable compared to bond-center and tetrahedral sites [41].

For most stable bond center site, the C-H bond length is 1.14 Å for both the nearest carbon atoms. The observed C-H bond length is good agreement with reported C-H bond length in range of ≈ 1.07 Å to 1.17 Å. In present calculation for fully relaxed structure, we observed higher dilation (68%) on C-C bond length. A broad range of dilatation for C-C bond is reported, with values ranging from 39 to 52% using different computational approaches [10]. The broad discrepancy for C-H bond length and C-C dilatation are reported in literature may be because of the use of very small carbon clusters or super cell and sometime relaxations of only first or second neighbor lattice.

In present case the BC site is 1.38 eV and 1.97 eV more stable than T and H site, respectively. Recently, DFT calculations using bigger size super cell found the BC site to be more stable [9, 10, 18, 42, 43]. We also computed H at anti-bond site but this structure was unstable and lattice relaxation push the H atom at tetrahedral site. The difference in energies is higher than the previously reported DFT calculation [9, 10, 18, 42, 43] because use of bigger supercell in present case. The comparison of some selected reports are given are made in Table I. The zero of total energy is taken to be the lowest energy structure in present study.

Fig. 3 depicted the density of states (DOS) of different used H atom site in diamond, (a) bond-center, (b) Tetrahedral and (c) Hexagonal. Hydrogen at bond center, tetrahedral and hexagonal sites give energy levels (Kohn-Sham level) in the band gap $\approx E_c$ -2.14 eV, $\approx E_c$-1.64 eV, and $\approx E_c$-0.64 eV, respectively, below the conduction band. The positions of KS levels shift toward the conduction band with increase of inter separation of H atom from the carbon atom along the <111> direction, which is consistent with the previous DFT calculations [41]. The formation of DOS hydrogen at ET site also reported at $\approx$ 0.5 eV above the middle of the energy gap emerges [11]. The formations of electronic DOS for H at these sites are caused by the formation of dangling bond by the relaxation and the breaking of the C-C bond.

**B. Hydrogen atom pairs in pure Diamond**

For hydrogen atom pairs, we have examined few possible structures. The relaxed geometry of the hydrogen pairs are shown in fig. 4, (a) $H_2^*$ complex, one H atom is at the C-C bond center and the second H is at the anti-bonding site behind the C atom along C-C axis, (b) $H_{2BC}$ complex, both the hydrogen at puckered bond center, (c) $H_{2TH}$ complex, one hydrogen at tetrahedral and second hydrogen at hexagonal site and (d) $H_{2TT}$ complex, both hydrogen at tetrahedral site. In our calculation, $H_2^*$ complex is found to be most stable within the computed structures. For the $H_2^*$ complex, the bond-centered hydrogen atom is no longer equidistant from the neighboring carbon atoms. The C-H bond lengths are 1.03 Å from one carbon and 1.35 Å from the other carbon atom and for anti-bonded H atom the C-H bond length is 1.03 Å. The calculated values are closed agreement with the previous reported values, 1.01 Å and 1.28 Å from its neighboring carbon atoms and for anti-bonding site, 1.01 Å, using hybrid DFT, cluster model and semiempirical electron delocalization molecular orbital theory [44, 45]. The structure predictions were not reported by the Goss et al [9], which make difficult to make comparison.

The $H_{2BC}$ complex is computed after considering two hydrogen atoms at tetrahedral site from the nearest carbon atom as a initial, to search the proposed stable structure by Saada et al [11]. Full relaxation give the structure as shown in fig. 4 (b) similar reported in reference [11], which is the second stable structure in our calculation. Here, both the hydrogen atoms have 1.03 Å bond lengths and angles are symmetric to each other. The calculated C-H bond length (1.03 Å) is similar than reported by the Saada et al [11]. Both the H atoms make 30 degree < HCC bond angles, with sharing the broken C-C bond. The inter separation between these two carbon atom is calculated to be 2.38 Å, which is higher than the C-C bond length (2.28 Å) when one H atom site at bond center site (fig. 2 (a). The distance of the hydrogen atom to the further of the carbon neighbours is 1.58 Å in present study. In case of Saada et al. [11], they reported this distance about 1.8 Å and an assignment was made with the estimated distance for the H1 EPR centre. They predicted ET site 2.5 eV lower in energy than the $H_2^*$ complex.

The $H_{2TH}$ complex is computed considering one hydrogen atom at T site and second at H site, full relaxed structure leads the both H atoms closed similar to hydrogen molecule. The inter atomic distance (0.67 Å) for two hydrogen is comparable with bond length (≈ 0.70 Å) in hydrogen molecule [46]. The next H pair complex ($H_{2TT}$) is computed with two hydrogen atoms are two tetrahedral sites. This complex is found more unstable in diamond. The details are given in Table II.

The electronic DOS for Hydrogen atom pairs are shown in fig. 5 (a) $H_2^*$, (b) $H_{2BC}$, (c) $H_{2TH}$ and (d) $H_{2TT}$ complex. For $H_2^*$ complex, no DOS are observed, which shows the electrical inert of this defect. The $H_{2BC}$ complex leads the creation of Kohn-sham (KS) level close to valance band at ≈ $E_\upsilon$ +0.22 eV, which is close agreement with previously reported KS levels [9]. The $H_{2TH}$ complex does not create KS level in band gap and $H_{2TT}$ complex leads two KS levels at ≈ 1.06 eV and ≈ 2.06 eV below the conduction band, inside the diamond band gap.

**IV. HYDROGEN IN BORON DOPED DIAMOND**

In case of isolated hydrogen in boron doped diamond, many efforts have been made to calculate the electronic properties and structure. We have also used these two different structures to investigate the electronic structure and stability of boron-hydrogen complexes. The corresponding relaxed structures are depicted in fig. 6 for the neutral defect. Our calculations reveal that hydrogen at the bond center site is only a saddle point, where as puckered bond center position have the lowest energy for these complex [17]. In case of puckered position, hydrogen is off the bond center B-C axis and makes a 106 degree angle which is less than BHC. It is also shown that distance between boron-hydrogen is larger than the hydrogen-carbon distance, but their difference reduced in puckered position. The values of bond length found for boron-hydrogen, hydrogen-carbon is agreed well with earlier results. Most of the previous works [17, 18-20, 48] with supercells and cluster approaches concluded that puckered position have higher stability for the isolated hydrogen in boron doped diamond. The puckered position allows relaxing the stress on the hydrogen-carbon and boron-hydrogen bonds. For the DOS of boron-hydrogen complexes corresponding to hydrogen at bond center and puckered positions are in reference [17]. This confirms that hydrogen at bond center position gives an energy level in the band gap of ≈ Ec-0.6 eV below the conduction band and there is no state found in the gap for the puckered position. The energy level $E_c$-0.6 eV for the on bond configuration appears as an independent band and not connected to conduction band as in earlier report [21] by using large supercell in this case. The details of the lowest energy sites and corresponding positions of the energy levels in the gap compared with previous studies are given in table III.

For the hydrogen atom pairs in boron doped diamond the different calculated configurations are described in fig. 7 for neutral B-$H_2$ complex. The previous researchers correspond to the lowest energy atomic structures determined in other ab initio calculations [18-20, 48]. We considered the case of B-$H_2^*$, first H atom is at the B-C bond center and the second H is at the antibonding site behind the B atom along B-C axis. In case of, B-$H_{2P}^*$ the H atoms are on bonding and antibonding puckered positions. In case of B-$H_{2BC}$, the one H atom at the B-C bond centered puckered position and the other one at the puckered position of one of the C-C bond. While in case of B-$H_{2CC}$, two H atoms are at the puckered bond center positions of two C-C bonds. It is found that the B-$H_{2BC}$ is the lowest energy structure for the neutral B-$H_2$ complex in agreement with reference [18] which opposes the earlier reports [20, 21]. Summary of these sites are given in table IV. The DOS of the relaxed B-$H_2$ complexes, for B-$H_2^*$, B-$H_{2P}^*$, B-$H_{2BC}$, and B-$H_{2CC}$ have been shown in reference [17]. The KS levels and the total energy are given in table II and compared to previously reported results. The DOS observed in the band gap for the most stable structure is consistent with the energy level observed in band structure calculations for neutral B-$H_2$ complex [49].

For multiple hydrogen atom trapping in B-doped diamond different possible structures are computed. Fig. 8 shows the relaxed geometry for three hydrogen atom in diamond cell, (a) B-H+$H_2^*$ complex, One hydrogen at pucker bond center close to boron and two hydrogen far from boron along <111> axis, one at C-C bond center and second at anti-bonding site, similar to $H_2^*$ complex, (b) B-$H_2$+H complex, two hydrogen atoms close to boron atom and one hydrogen far from B at C-C bond center site, (c) B+$H_2^*$+H complex, all the three H atoms far from the B atom, one hydrogen at C-C bond center, second H at anti-bonding site and third at Hexagonal site, and (d) B-H+$H_2$ complex, one hydrogen at puckered bond center site close to B and two H far from the B, one Hydrogen at one tetragonal and second at second tetragonal site in same section.

In case of B-H+$H_2^*$ complex, H atom at puckered position has 1.20 Å B-H, 1.16 Å H-C bond length and create 103 degree < BHC angle. The two hydrogen far from the B-H complex make $H_2^*$ structure with equal bond length (1.03 Å). In B-$H_2$+H complex, two hydrogen stay close to boron with the similar C-H bond length (1.03 Å). The one atom far from boron atom stays on bond center and creates 1.12 Å C-H and 1.09 Å H-C bond lengths. The B+$H_2^*$+H complex, two hydrogen atom in bonding and anti-bonding site far from the boron atom from a $H_2^*$ structure with a hydrogen at hexagonal site. The C-H bond lengths (1.03 Å) of $H_2^*$ structure is exactly same as we studied in fig. 4 (a) and in fig. 8 (a). For B-H+$H_2$ complex, one H atom stays at puckered bond center site, which has 1.20 Å B-H, 1.15 Å H-C bond lengths and 105 degree (<BHC) angle. Two atoms far from the boron atom male hydrogen molecular structure with the 0.67 Å bond length, similarly observed in fig. 4 (c). It is pointed out that H atom at puckered position has slightly different bond lengths and bond angles for B-H+$H_2^*$ and B-H+$H_2$ complex.

These calculations suggest the B-H+$H_2^*$ complex structure is most stable within the computed structures for the B-$H_3$ complex. Here, it is very difficult to make the comparison as B-$H_3$ complexes in diamond were not reported in the literature. Investigation of bigger complexes makes its exhaustive with computational expanse. Several other possible structures also studied but all are found to be highly unstable and not presented here.

## V. CONCLUSION

The present work reports electronic structure and stability of of H related defects in pure and boron doped diamond. Detailed calculations revealed that isolated hydrogen defect is stable at bond center sites for pure diamond and bond center puckered site for boron doped diamond. For hydrogen atom pairs, $H_2^*$ defect is most stable for pure diamond. For boron doped diamond B-$H_{2BC}$ complex (one H atom at the B-C bond centered puckered position and the other one at the puckered position of one of the C-C bond first neighbor of the B atom) is most stable. We discussed our results and detailed comparisons made with previous calculated results. Current review of these theoretical calculations of electronic structure and stability of H defects in pure and boron doped diamond will be valuable for researchers to explore the tremendous properties of diamond in future.

## VI. ACKNOWLEDGEMENTS

A. K. gratefully acknowledges the financial support from Agence Nationale pour la Recherche (ANR 06 BLAN 0339-02), France and Department of Science and Technology (DST), India. We thank to Prof. Julien Pernot and Prof. Laurence Magaud, Neel Institute Grenoble, France for enlightening discussions. The Xcrysden program is used to plot the relaxed position of atoms [50].

**FIGURE CAPTIONS:**

FIG.1. Schematics of the possible hydrogen sites in diamond cell. Carbon atoms are shown as golden circles and hydrogen atoms by black circle. The Tetrahedral (T) site lies equidistant from four carbon atoms and the hexagonal (H) site lies midway between two T sites. The bond-center (BC) site is the midway between two carbon atom sites and anti-bonded (AB) sites is opposite the bond center site along the C-C axis as showing in the figure.

FIG.2. The relaxed structures of isolated Hydrogen atom at, (a) Hydrogen at Bond center site, (b) Hydrogen at Tetrahedral site and (c) Hydrogen at Hexagonal site.

FIG.3. Density of states of isolated Hydrogen at, (a) Hydrogen at Bond center site, (b) Hydrogen at Tetrahedral site and (c) Hydrogen at Hexagonal site.

FIG.4. The relaxed structures of hydrogen pairs, (a) $H_2^*$ complex, one Hydrogen at bond center and other at anti-bonding site, (b) $H_{2ET}$ both the hydrogen at equilateral triangle sites, (c) $H_{2TH}$, one hydrogen at tetragonal site and other at hexagonal site, and (d) $H_{2HH}$ complex, one hydrogen at hexagonal site and second hydrogen at other hexagonal site in diamond cell.

FIG.5. The density of states of hydrogen pairs, a) $H_2^*$ complex, (b) $H_{2ET}$ complex, (c) $H_{2TH}$ complex, and (d) $H_{2HH}$ complex.

FIG.6. The relaxed structures of B-H complex, (a) H along B-C axis and (b) H at puckered position.

FIG.7. The relaxed structures of B-$H_2$ complexes, (a) B-$H_2^*$, (b) B-$H_{2P}^*$, (c) B-$H_{2BC}$, and (d) B-$H_{2CC}$

FIG.8. The relaxed structures of three hydrogen atoms in B-doped diamond matrix, (a) Hydrogen along the B-C axis and (b) Hydrogen at puckered position.

TABLE I: Summary and comparison of theoretical work on Hydrogen in diamond matrix. The zero of total energy (eV) is taken to be the lowest energy.

| Reference | Method | Bond center | Sites Tetrahedral | Hexagonal |
|---|---|---|---|---|
| Our work | GGA-DFT, 216-512 C atom unit cell | 0.0 | 1.38 | 1.97 |
| [9, 18] | LDA-DFT, 64-216 C atom unit cell | 0.0 | 1.0 | 1.7 |
| [41] | LDA-DFT, 64 C atom supercell | 0.0 | 0.95 | 1.52 |
| [11] | TB calculations 216 C atom supercell | 0.0 | 0.5 | … |
| [42] | PIMD simulations 64 C atom supercell | 0.0 | 1.44 | … |
| [8] | LDA-DFT $C_{26}H_{30}$ cluster | 0.0 | 1.9 | … |
| [7] | PRDDO approximation $C_{26}H_{30}$ cluster | 0.0 | 2.7 | … |
| [36] | HF approximation $C_{26}H_{30}$ cluster | … | 0.0 | 0.83 |

TABLE II: Summary and comparison of theoretical work on Hydrogen pairs in diamond matrix. The zero of total energy (eV) is taken to be the lowest energy.

| Reference | Method | Structure | | | |
|---|---|---|---|---|---|
| | | $H_2^*$ | $H_{2BC}$ | $H_{2TH}$ | $H_{2HH}$ |
| Our work | GGA-DFT, 216-512 C atom unit cell | 0.0 | 0.73 | 2.22 | 4.97 |
| [9] | LDA-DFT, 64 C atom unit cell | 0.0 | 0.80 | 2.1 | … |
| [44] | Hybrid DFT calculations $C_{44}H_{42}$ cluster | 0.0 | 0.66 | … | … |
| [8] | LDA-DFT, $C_{26}H_{30}$ cluster | 0.0 | 3.32 | … | … |
| [47] | HF approximation $C_{26}H_{30}$ cluster | 0.0 | 2.23 | … | … |
| [11] | TB calculations 216 C atom supercell | 2.5 | 0.0 | … | … |

TABLE III: Summary and comparison of theoretical work on B-H complexes in diamond matrix. The zero of total energy (eV) is taken to be the lowest energy.

| Reference | Method | Structure | |
|---|---|---|---|
| | | H at puckered position (Ks level) | H at bond-centered position (KS level) |
| Our work | GGA-DFT, 216-512 C supercell | 0.0 (No level) | 0.3 ($E_C$-0.6 eV) |
| [18, 48] | LDA-DFT, 64 C atom supercell | 0.0 (No level) | 0.6 |
| [19] | LDA-DFT 123-165 clusters | (No level) | … |
| [20] | DFT cluster, 35 C cluster | (level in gap) | … |
| [21] | GGA-VASP, 64 C atom unit cell | ($E_C$-1.0 eV) | … |

● KS - denotes the Kohn-Sham levels.

TABLE IV: Summary and comparison of theoretical work on B-H$_2$ complexes in diamond matrix. The zero of total energy (eV) is taken to be the lowest energy.

| Reference | Method | KS level | Relative total energy |
|---|---|---|---|
| (B – H$_{2BC}$ complex) | | | |
| Our work | GGA-DFT, 216-512 C supercell | $E_v$+1.29 eV | 0 |
| [49] | LDA-DFT, 64-216 C atom supercell | $E_v$+1.2-1.3 eV | |
| (B – H$_{2P}^*$ complex) | | | |
| Our work | GGA-DFT, 216-512 C supercell | $E_v$+1.3 eV | 0.77 |
| [21] | GGA-DFT, 64 C atom unit cell | Not shown | |
| (B – H$_2^*$ complex) | | | |
| Our work | GGA-DFT, 216-512 C atom unit cell | $E_v$+0.87 eV | 0.88 |
| (B – H$_{2CC}$ complex) | | | |
| Our work | GGA-DFT, 216-512 C atom supercell | $E_v$+1.52 eV $E_v$+2.04 eV | 1.95 |
| [20] | DFT cluster, 35 C atom cluster | n-type conductivity | |

● KS - denotes the Kohn-Sham levels.

**Fig. 1.**

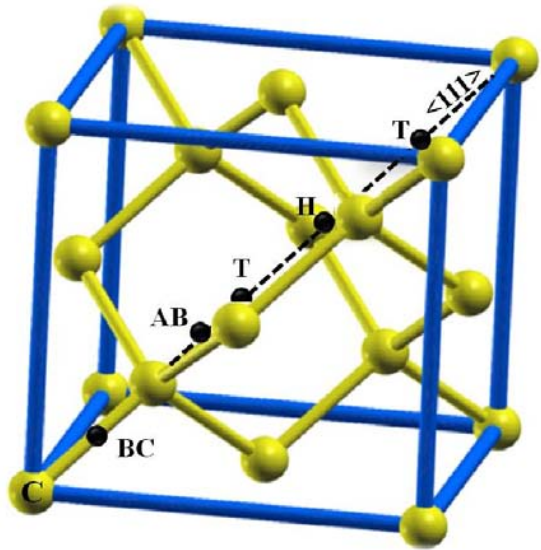

**Fig.2.**

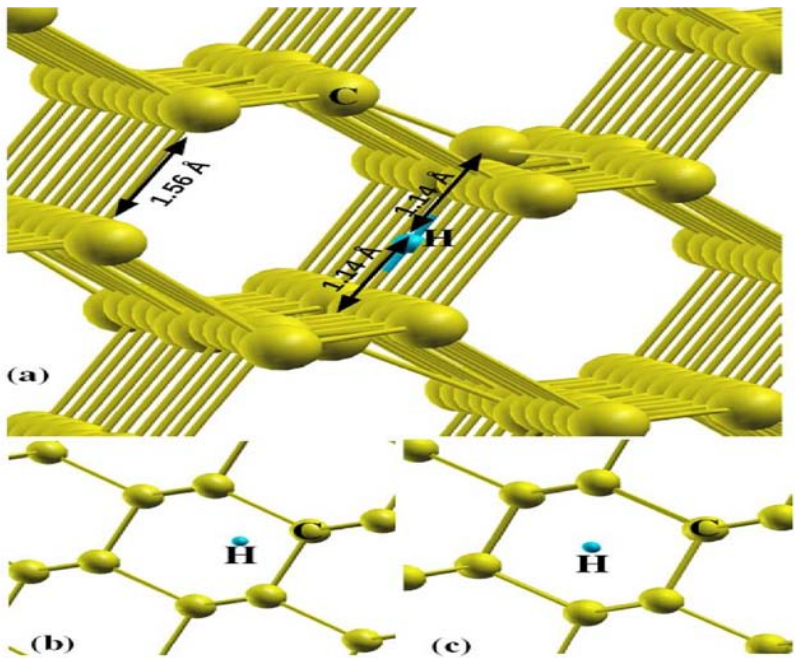

**Fig.3.**

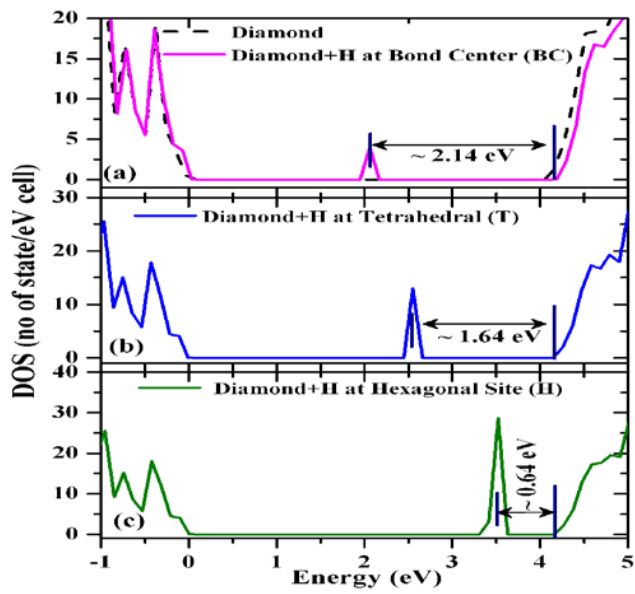

**Fig.4.**

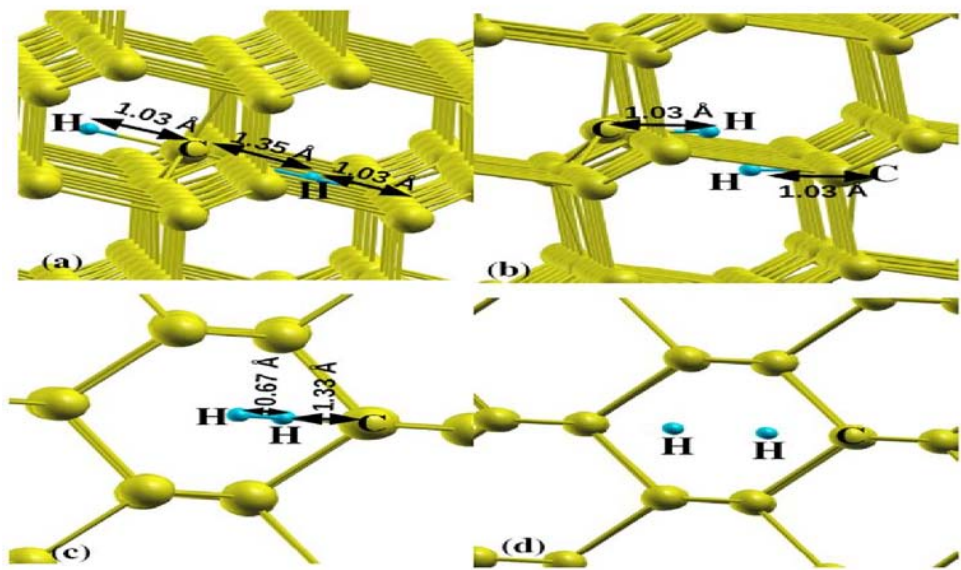

**Fig.5.**

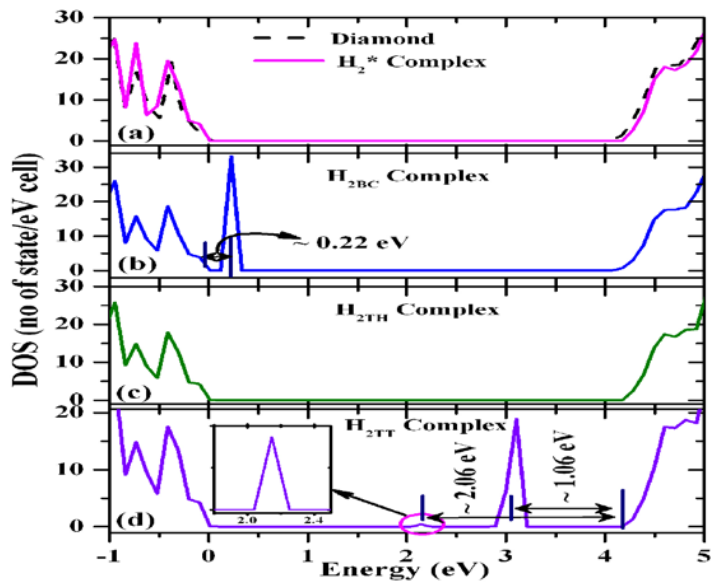

**Fig.6.**

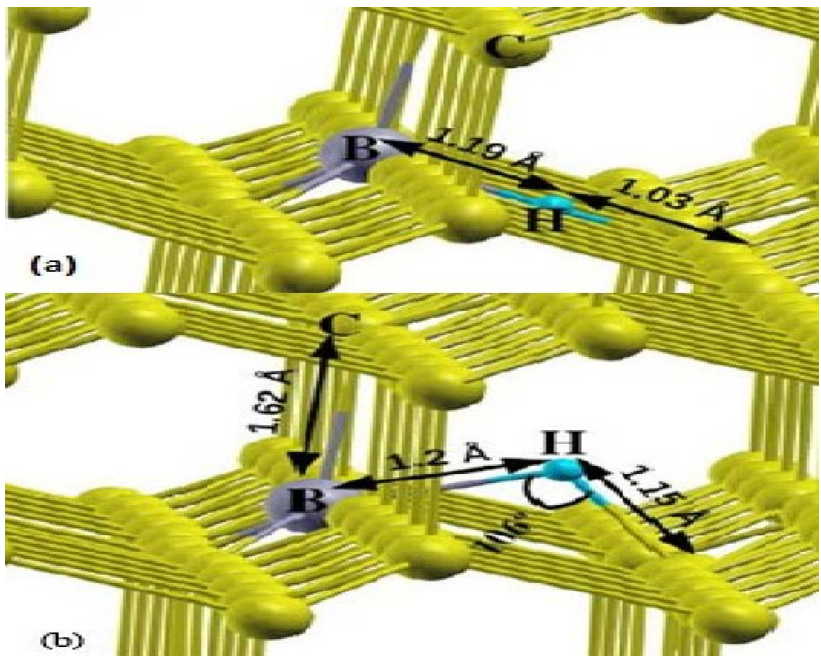

**Fig.7.**

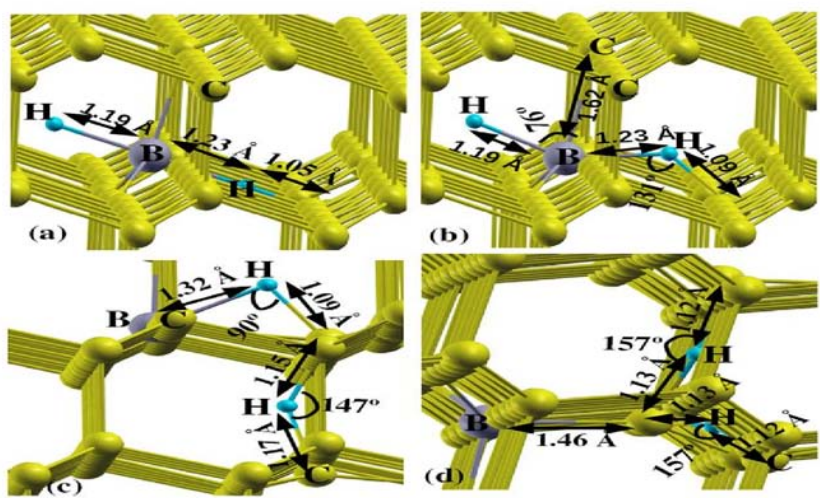

**Fig.8.**

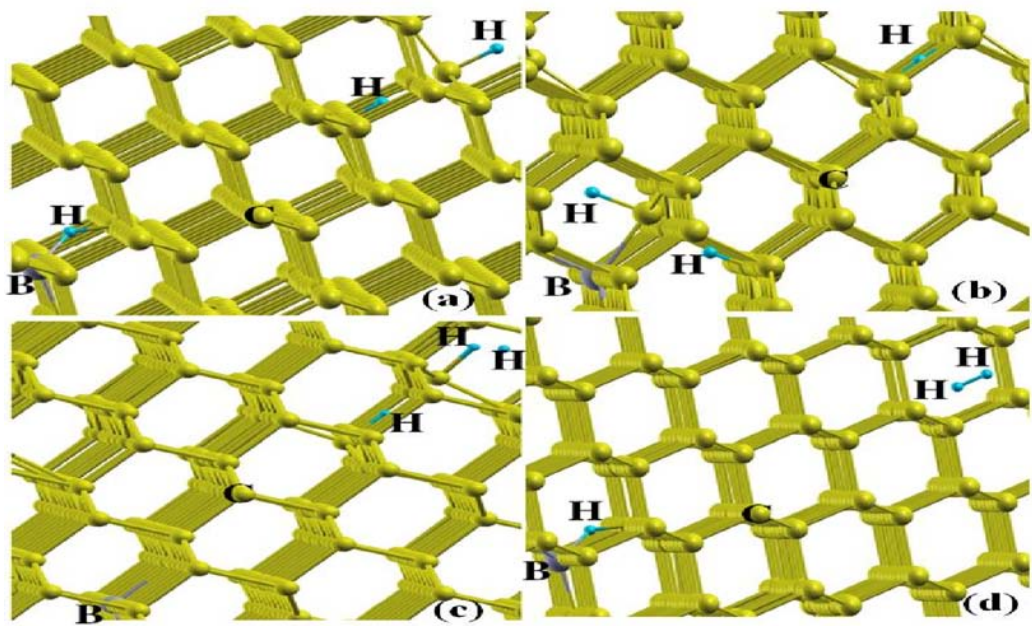

22